\documentclass[superscriptaddress,aps,amsmath,amssymb,twocolumn,showpacs,pre]{revtex4-1}
\usepackage{graphicx,color}
\usepackage{etoolbox}
\graphicspath{{./}{./figures/}}
\usepackage{dcolumn}
\usepackage{bm}
\usepackage[percent]{overpic}
\begin{document}
\bibliographystyle{unsrt}

\title{Strong pinning of propagation fronts in adverse flow}
\author{Thomas Gueudr\'e}
 \affiliation{CNRS-LPT, Ecole Normale Sup{\'e}rieure 75231 Cedex 05 Paris, France} 
\author{Awadhesh Kumar Dubey}
\affiliation{Universit\'e Paris-Sud, CNRS, Laboratoire FAST, UMR 7608, Orsay F-91405, France.}
\author{Laurent Talon}
\affiliation{Universit\'e Paris-Sud, CNRS, Laboratoire FAST, UMR 7608, Orsay F-91405, France.}
\author{Alberto Rosso}
\affiliation{Universit\'e Paris-Sud, CNRS, LPTMS, UMR 8626, Orsay F-91405, France.}

\date{\today\ -- \jobname}


\begin{abstract}
Reaction fronts evolving in a porous medium exhibit a rich dynamical behaviour. In presence of an adverse flow, experiments show that the front slows down and eventually gets pinned, displaying a particular sawtooth shape. Extensive numerical simulations of the hydrodynamic equations confirm the experimental observations. Here we propose a stylized model, predicting two possible outcomes of the experiments for large adverse flow: either the front develops a sawtooth shape, or it acquires a complicated structure with islands and overhangs. A simple criterion allows to distinguish between the two scenarios and its validity is reproduced by direct hydrodynamical simulations. Our model gives a better understanding of the transition and is relevant in a variety of domains, when the pinning regime
is strong and only relies on a small number of sites.
\end{abstract}
\maketitle

In the systems separated in distinct phases, the dynamics is controlled by the behaviour of the propagating fronts. Those fronts pervade a broad variety of domains in physics, ranging from chemotaxis \cite{adler_chemotaxis_1966} and plasma physics \cite{boyd2003physics} to flames front \cite{jarosinski_combustion_2009} or epidemics, therefore triggering much activity in their modelling (for a recent review, see \cite{fort_progress_2008}). One of the cornerstones in this field is the celebrated Fisher-KPP theory, describing the front propagation in reaction-diffusion systems \cite{fisher_wave_1937}.  However, this approach was limited to systems with no advection, i.e. not undergoing any fluid flow, despite its physical importance. Coherent fluid-like motion strongly impacts the dynamics of the fronts \cite{edwards_poiseuille_2002} and remains a challenging problem, whether because of the appearance of turbulence \cite{schwartz_chemical_2008}, or because of the influence of a disordered media \cite{xin_front_2000,koptyug_advection_2008}. One natural disordered environment for propagation fronts is a porous medium. Some examples were investigated in the petrol industry and aeronautics with attempts to address the evolution of a flame front in a gas filter \cite{korzhavin_dynamics_1997,kuo_theory_1973}. Recently, experiments on self-sustained chemical reactions have allowed a fine and controlled examination of the propagation fronts in porous medium, revealing some striking features by direct observation  \cite{atis_self-sustained_2012,atis_autocatalytic_2013}. 

\begin{figure}
\includegraphics[width=0.45 \textwidth]{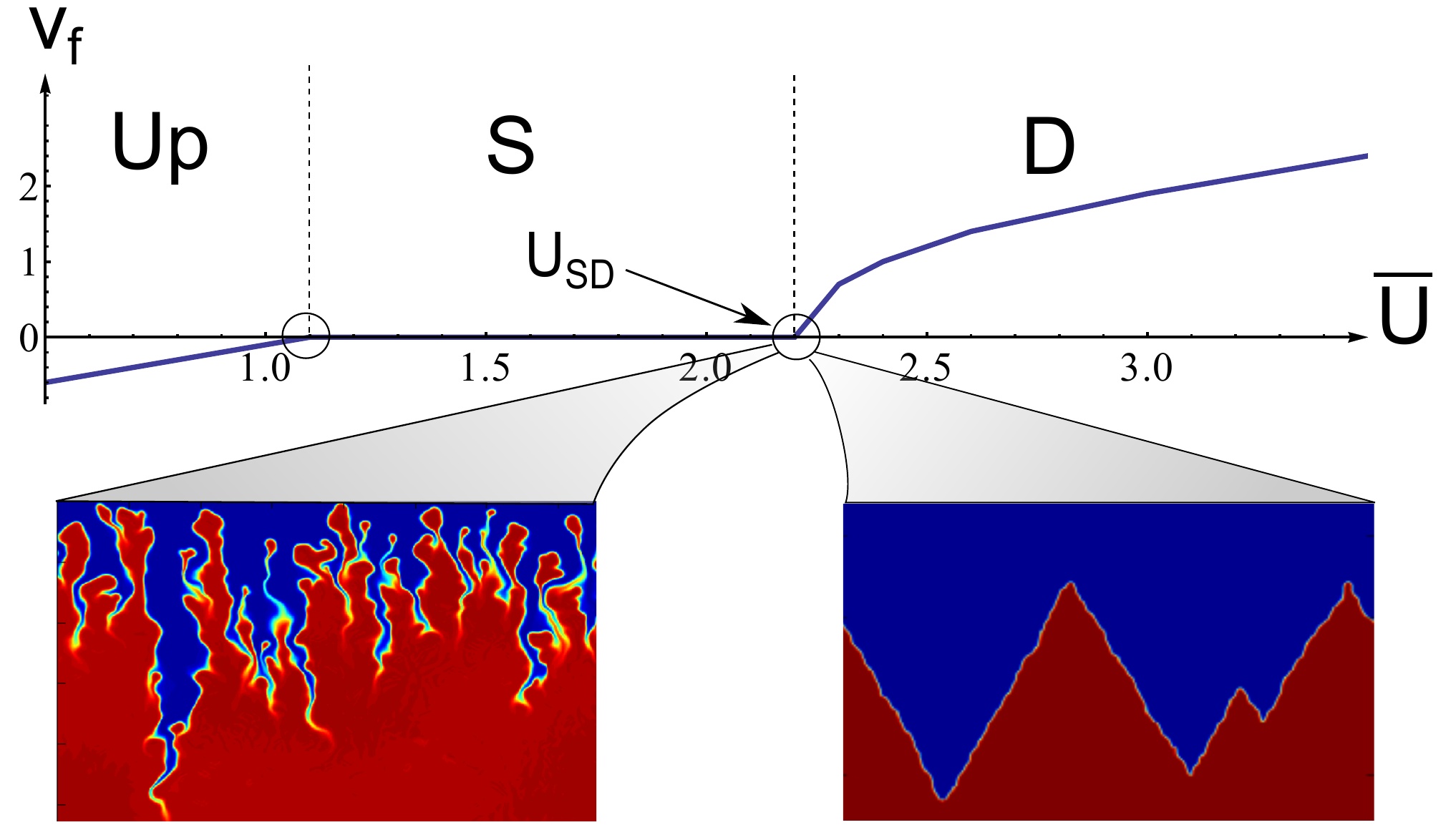}
\centering
\caption{Average speed of the front $V_f$ as a function of the injection speed $\overline{U}$ (the convention chosen is $\overline{U}>0$ for an flow from top to bottom of the cell). Depending on the sign of $V_f$, different regimes $\text{Up}$, $S$ and $D$ are defined. For $\overline{U}=U_{SD}$, we observe a transition between a static front ($V_f =0$) and a downstream front ($V_f >0$). {\em Bottom part}: Hydrodynamical simulations of the front, for $\overline{U} \sim U_{SD}$ and for different permeability distributions. Two scenarios are observed: a regular saw-tooth shape (right) or a complicated shape with overhangs (left).}
\label{numericsphase}
\end{figure}

The experimental setup employs an autocatalytic reaction invading a cell filled with a solution of reactants. To reproduce porosity, the cell also contains a mixture of glass spheres of different sizes. The reaction starts at the bottom of the cell and, in the absence of advection flow, develops into an almost flat front propagating upwards with constant chemical speed $|V_{\chi}| =\sqrt{D_m \alpha/2}$ and width $\ell_\chi =D_m /|V_\chi|$, $D_m$ being the molecular diffusion constant and $\alpha$ the reaction rate. In presence of an adverse flow injected from the top at speed $\overline{U}$, the porosity generates a fixed random velocity map of the fluid with short range correlations of characteristic length $\ell_d$. A rich dynamical phase diagram is observed as a function of the flow velocity $\overline{U}$, the control parameter of the experiment (see Fig.\ref{numericsphase}). In particular, the self-sustained fronts can travel downstream along the flow ($D$), remain static over a range of flow rate values ($S$) or move upstream ($\text{Up}$). In all these phases, the heterogeneities make the front rough and the dynamics proceed by random jumps called avalanches displaying a free scale statistics.

Here we focus on the transition between the static and the downstream regimes, occuring at the threshold $U_{\text{SD}}$ (see Fig.\ref{numericsphase}). Hydrodynamical simulations show two different scenarios: either the invading chemical reaction is completely washed away for $\overline{U} >U_{\text{SD}}$, or some stagnant chemicals remain trapped in the porous media for any $\overline{U}$. In the first case, approaching $U_{\text{SD}}$ from below, the front is largely deformed into a sawteeth structure (see Fig.\ref{numericsphase} bottom right), while in the second case, the interface adopts a very rough and complicated structure with overhangs (see Fig.\ref{numericsphase} bottom left). Experiments typically correspond to the first scenario but the second one has also been observed in very contaminated cells \cite{atis_self-sustained_2012}.

In this paper, we describe the front propagation with a stylized model controlled by two parameters that can be easily measured in experiments. This model gives a simple criterion to discriminate between both scenarios, depending only on the behaviour of the disorder distribution close to $0$. The critical threshold $U_{\text{SD}}$ and the shape of the front can be characterised. Comparison with hydrodynamical {\em ab initio} simulations using the Darcy equation shows a perfect agreement with our results. Although our approach addresses questions raised by the experiments of \cite{atis_self-sustained_2012}, the results of this stylized model are much more general and relevant to all systems where the transition between a static and a moving regime is controlled by a small number of pinning sites \cite{yang_nanostructured_1997,maiorov_synergetic_2009,koshelev_theory_2011,dalmas_crack_2009}.

{\em From first principle hydrodynamics to a simple statistical model.} The flow field $\vec{U}(\vec r)$ can be computed via the Darcy-Brinkman equation:
\begin{align}
&\vec{\nabla} \cdot \vec{U}(\vec r) = 0 \\
&\vec U (\vec r) = -\frac{K(\vec r)}{\eta}  \vec{\nabla} P + K(\vec r) \Delta \vec U
\end{align}
where $P(\vec r)$ is the pressure field, $\eta$ the fluid viscosity and $K(\vec r)$ the local permeability. Due to the incompressibility, the mean fluid velocity is fixed to the injection rate $\overline{U}$. Once the hydrodynamic problem is solved, the concentration of the chemicals $C(\vec r, t)$ obeys an advection-diffusion equation (see \cite{jarrige10a}):
\begin{align}
\frac{\partial C}{\partial t}  + \vec U.\vec \nabla C = D_m \Delta C + \alpha C^2 (1-C)
\end{align}
The effect of the disorder is incorporated in the permeability $K(\vec r)$, usually modelled as a random field, correlated over a distance $\ell _d$. Here we study the front geometry for different permeability distributions: the log normal distribution, often employed to model permeability \cite{matheron}, and the distributions belonging to the Weibull family of parameter $\delta$. On Fig.\ref{numericsphase} are displayed typical fronts for both log-normal distributed (bottom left) and Weibull distributed (bottom right, with $\delta = 0.8$) permeability fields. Those are generated using a standard method detailed in the Appendix A. Both $U(\vec r)$ and $C(\vec r,t)$ were solved using a Lattice Boltzmann scheme (see \cite{talon03, ginzburg08c}). We ran the simulations on a square grid of size $L=512$, up to $N=2000$ realisations.

\begin{figure}
\includegraphics[trim=11.5cm 8.5cm 1cm 1.5cm , width=0.45 \textwidth]{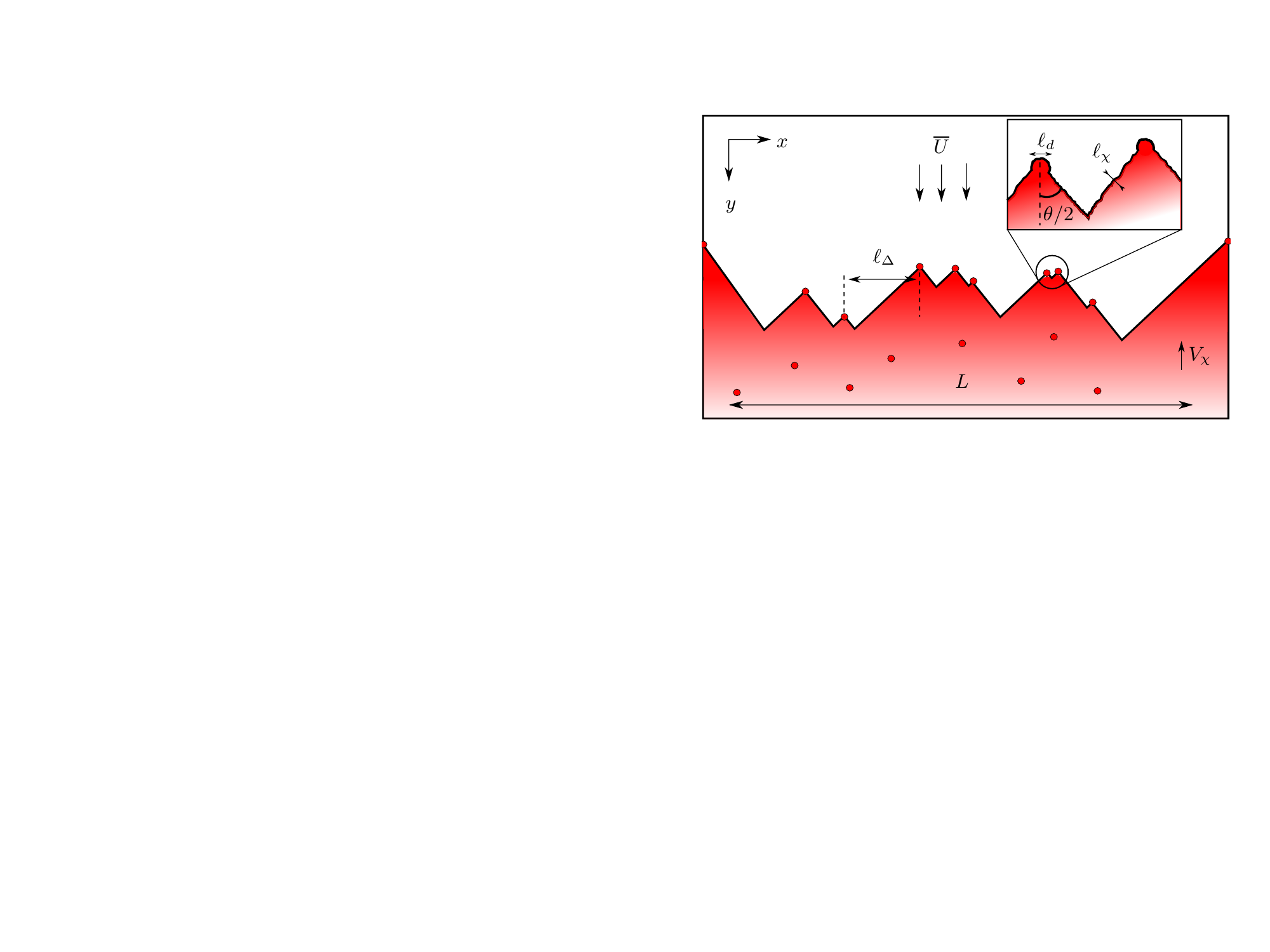}
\centering
\caption{Sketch of the stylized model: in the thin Eikonal limit $\ell_d \gg \ell _{\chi}$, the system can be described as a propagating front. Close to $U_{SD}$, the density of stagnant regions becomes small and the interface adopts a \textit{sawtooth} structure.}
\label{cartoon}
\end{figure}

In the experimental conditions $\ell_d \gg \ell_\chi$, the front lays in the so-called {\em thin front Eikonal} limit \cite{lecontemartinsalin,williamsconb}. In this limit, at each point of the front, the normal component of the interface velocity satisfies $\vec V_f (\vec{r}) \cdot \vec{n}= V_{\chi}+ \vec{U}(\vec{r})\cdot \vec{n}+ D_m \kappa $, where $\vec n$ is the unit normal vector and $\kappa$ the curvature of the front. For $\overline{U} \sim U_{SD}$, $\vec U(\vec r)$ is mainly directed along the $y$-axis $\vec U(\vec r) \sim (0,U(\vec r))$. It is natural to assume that $U(\vec r)$ is constant on patches of area $\ell_d^2$ and decorrelated between patches. The velocity of each patch is an independent random variable of average $\overline{U}$, distributed as:
\begin{align}
P_{\overline{U}}(U)=\frac{1}{\overline{U}} \phi(U/\overline{U})
\end{align}
where the scaling function $\phi(v)$ is independent of $\overline{U}$.
When $\overline{U} \leq U_{\text{SD}}$ the front is pinned by the very few stagnant sites where  $U( \vec r) <  |V_\chi|$. Hence the distance, $\ell_\Delta$ between them is much larger than $\ell_d$. In the neighbourhood of a pinning site, the front has a \textit{sawtooth} shape of angle $\theta$ and the front displays sawtooth-like structure (see Fig.\ref{cartoon}). $\theta$ can be computed observing that, in that regime, $\kappa \simeq 0$, $V_f(\vec r )=0$ and $U(\vec r ) \simeq \overline{U}$, thus the eikonal equation becomes \cite{atis_self-sustained_2012}:
\begin{align}
V_{\chi} + \overline{U} \sin (\theta /2)=0
\end{align}
Therefore the geometry of the frozen fronts is completely determined by the velocity-dependent angle $\theta$ and by the positions of the pinning sites. In particular the probability that a given patch of area $\ell_d^2$ is a pinning site is:
\begin{align}
\lambda &= \int _0^{|V_{\chi}|} P_{\overline{U}}(U) dU =\int _0^{|V_{\chi}|/\overline{U}} \phi(v) dv 
\label{lambdadef}
\end{align}
For large downstream injection rate $\overline{U} \gg |V_\chi|$, the value of $\lambda$ is controlled  only by the behaviour of $\phi(v)$ for $v$ close to $0$. In \cite{atis_self-sustained_2012} it was observed that the fluid velocity vanishes near the wall. To mimic that fact, we always set the interface pinned at the points $\vec r =(0, 0)$ and $\vec r =(L, 0)$. Therefore, if no stagnant patch inside the cell pins the front, except at the walls, the interface acquires a V shape that we call the \textit{depinned state}.

%

\begin{figure}
\begin{overpic}[width=0.50 \textwidth , height=55mm]{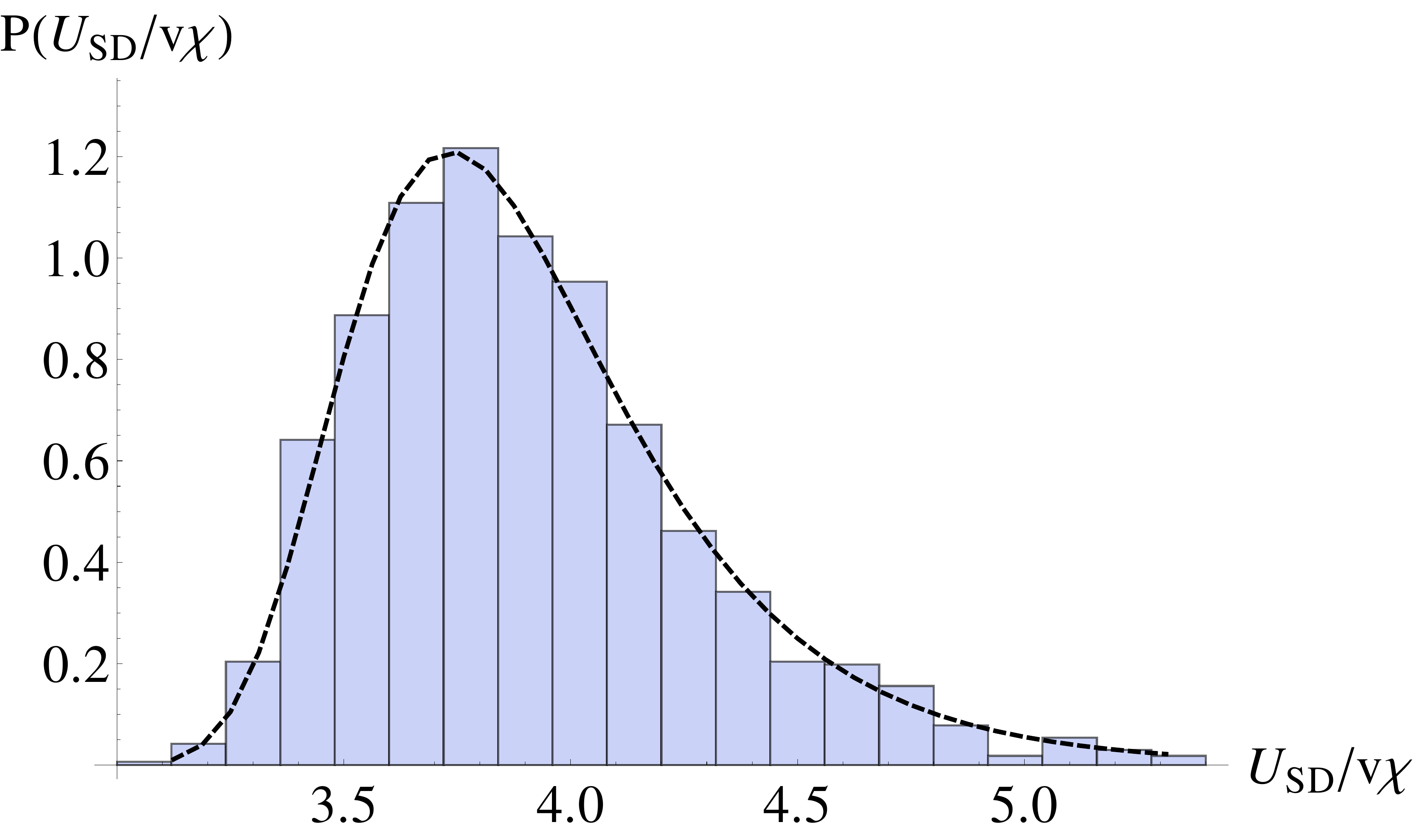}
\put(50,25){\includegraphics[scale=0.27]{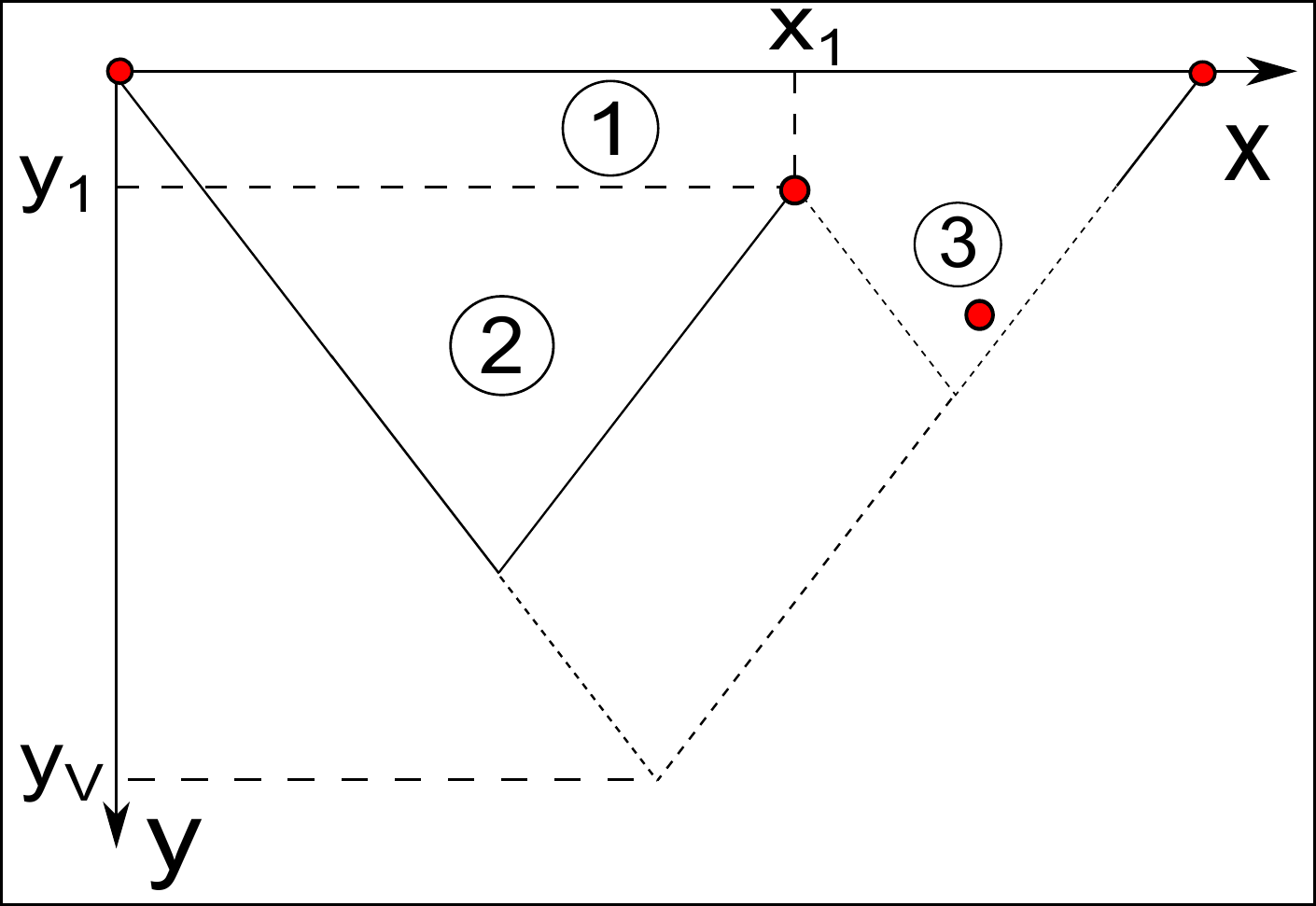}}
\centering
\end{overpic}	
\caption{Probability distribution of the velocity threshold $U_{SD}$. The histogram corresponds to the hydrodynamical simulations of $N=2000$ samples with a log-normal permeability, setting $L=512$, $V_{\chi}=0.0016$, $\overline{U} = 0.0036$ and $\ell _d =5.0$. The dashed line corresponds to the prediction of the stylized model (Eq.(\ref{Qdef})) for a log-normal $\phi (v)$ with a scale parameter $\sigma =0.315$ (see main text). Inset: sketch of the algorithmic recursive procedure.}
\label{fig:fitUSD}
\end{figure}

A central quantity for our analysis is $Q(y)$, the probability that, from $y=0$ to $y$, no stagnant patch is encountered. $Q(y)$ obeys to the differential equation:
\begin{align}
Q(y+dy)=\left(1-\lambda dy \left(L-2 \, \tan(\theta/2)\, y \right)/\ell_d^2 \right) Q(y)
\end{align} 
because the probability that no pinning occurs between $y$ and $y+dy$ is $1-\lambda dy (L-2 \, \tan(\theta/2)\, y)/\ell_d^2$  in an interval of size $L-2 \, \tan(\theta/2)\, y$. Hence:
\begin{align}
Q(y)=e^{-\lambda (L y + \, \tan(\theta/2)\,  y^2)/\ell_d^2}
\label{probdepinning}
\end{align}
This formula is valid up to  $y_V=\frac{L}{2  \, \tan(\theta/2)\, }$, value above which the front is in the depinned state. This quantity allows to introduce an efficient algorithm to generate the sites pinning the front: note $\epsilon$ a random number uniformly distributed in $(0,1)$, if $\epsilon> Q(y_V)$ the algorithm terminates with a V shape, while if $\epsilon< Q(y_V)$ the height of the first pinning site is $y_1=Q^{-1}(\epsilon)$ and its position $x_1$ is chosen at random in the segment of length $L-2 \tan(\theta/2 ) y_1$. This patch divides the segment into two pieces (see inset of Fig.\ref{fig:fitUSD}) and we recursively apply the algorithm on both pieces until no more stagnant patch is found.

Moreover, Eq.(\ref{probdepinning}) determines the statistics of the threshold $U_{SD}$. The probability of being in the depinned state for a certain injection rate $\overline{U}$ ($y=y_V$)  is given by:
\begin{align}
Q^{\text{dep}}_{\overline{U}}= \exp \left(-\frac{L^2 \int _0^{|V_{\chi}|/\overline{U}} \phi(v) dv }{\ell_d^2 \tan(\arcsin(|V_{\chi}|/\overline{U}))} \right)
\label{Qdef}
\end{align}
We note that this probability goes to $0$ quadratically in $L$. More generally, in $d$ dimensions, $Q(y_V)$ would decay as $\exp(-L^{-d})$. Hence the effect of the cell size on the transition is very strong and explains why washing a propagating front in disordered medium can be surprisingly hard. With raising $\overline{U}$, Eq.(\ref{Qdef}) exhibits two competing effects: the stagnant patches get decimated while the reaction front stretches (namely $\theta \to 0$) and explores a larger region. Assuming $\phi(v) \sim v^{\delta-1}$ when $v \to 0$, we get:
\begin{align}
Q^{\text{dep}}_{\overline{U}}= \exp \left(-\frac{L^2}{\ell_d^2} \left( \frac{|V_\chi|}{\overline{U}} \right)^{\delta-1}  \right)     \;     \text{  when   }  \frac{|V_{\chi}|}{\overline{U}} \to 0
\end{align}

The two scenarios pictured in Fig.\ref{numericsphase} now emerge naturally: if $\delta > 1$, the number of teeth decreases with $\overline{U}$ and the interface always gets depinned, while if $\delta < 1$, pinning sites proliferate and the front becomes rougher and rougher. The transition between the two regimes occurs at a critical value $\delta _c =1$: in that marginal case, the number of teeth remains constant. This prediction is well supported by the hydrodynamical simulations of the porous media for different $P_{\overline{U}}(v)$, where a clear transition towards roughening for $\delta <1$ is observed. In the experiments, the measured velocity map was fitted to a Log-normal distribution, decaying to $0$ as $v^{-1} \exp(-\log(v)^2) $, faster than the critical case, but not much. Hence, depinning indeed occurs. Note that the threshold speed $U_{SD}$ is itself random and depends on the realisation of the disorder. Its probability distribution $P(U_{SD})= \partial _{\overline{U}}Q^{\text{dep}}_{\overline{U}} |_{\overline{U}=U_{SD}}$ depends on the scaled velocity distribution $\phi(v)$. In  Fig.\ref{fig:fitUSD} we test the prediction of our model against hydrodynamical simulations for a log-normal permeability.  We assumed that the velocity of the fluid displays as well a log-normal distribution  $\phi(v)=(\sqrt{2 \pi } v \sigma)^{-1} e^{- (\log v)^2 / 2\sigma ^2}$ with a scale parameter $\sigma=0.315$ obtained from direct fit of the velocity distribution close to zero. A very good agreement \textit{with no adjusting parameter} is observed.

\begin{figure}
\begin{overpic}[width=0.47 \textwidth]{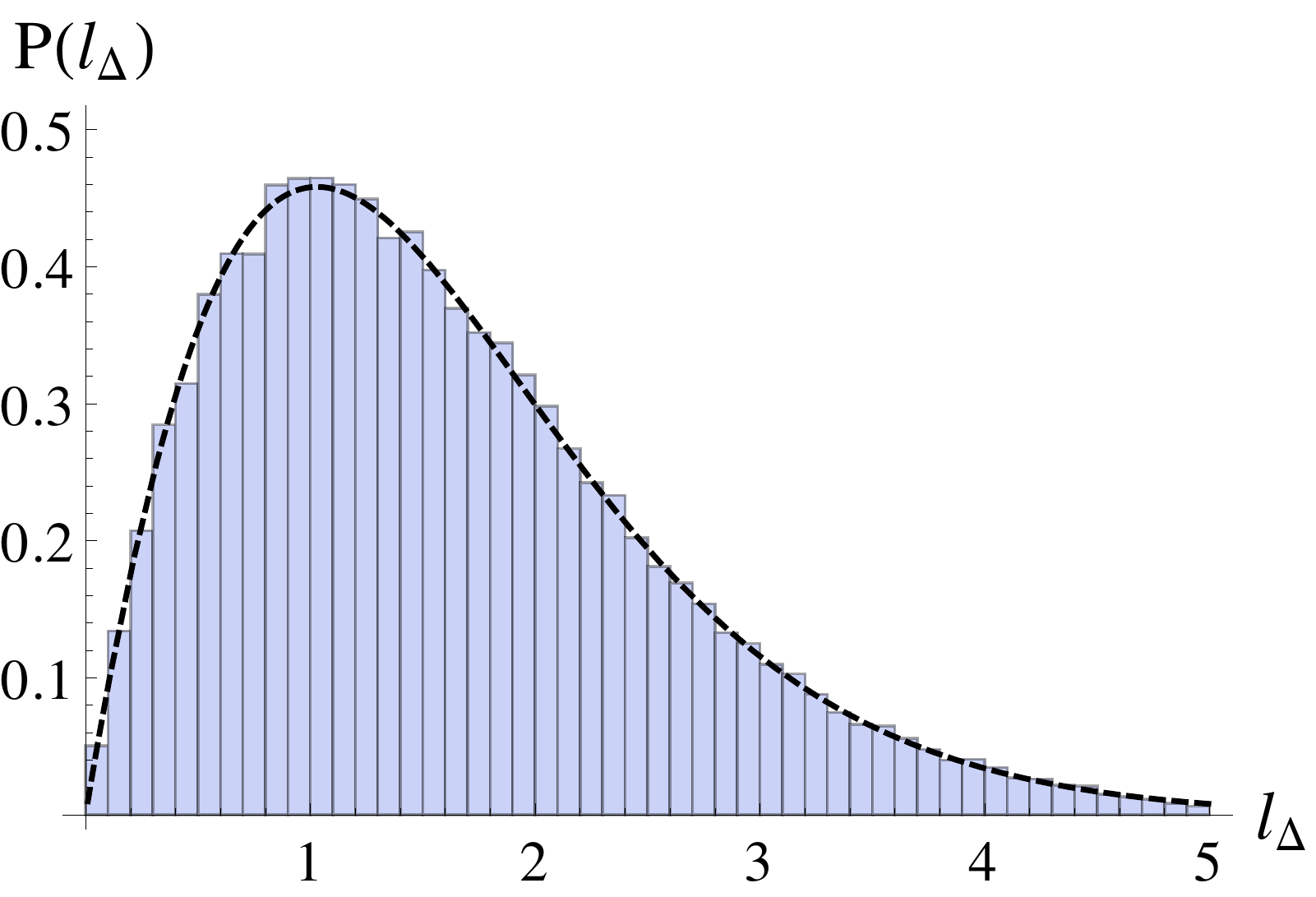}
\put(50,35){\includegraphics[scale=0.35]{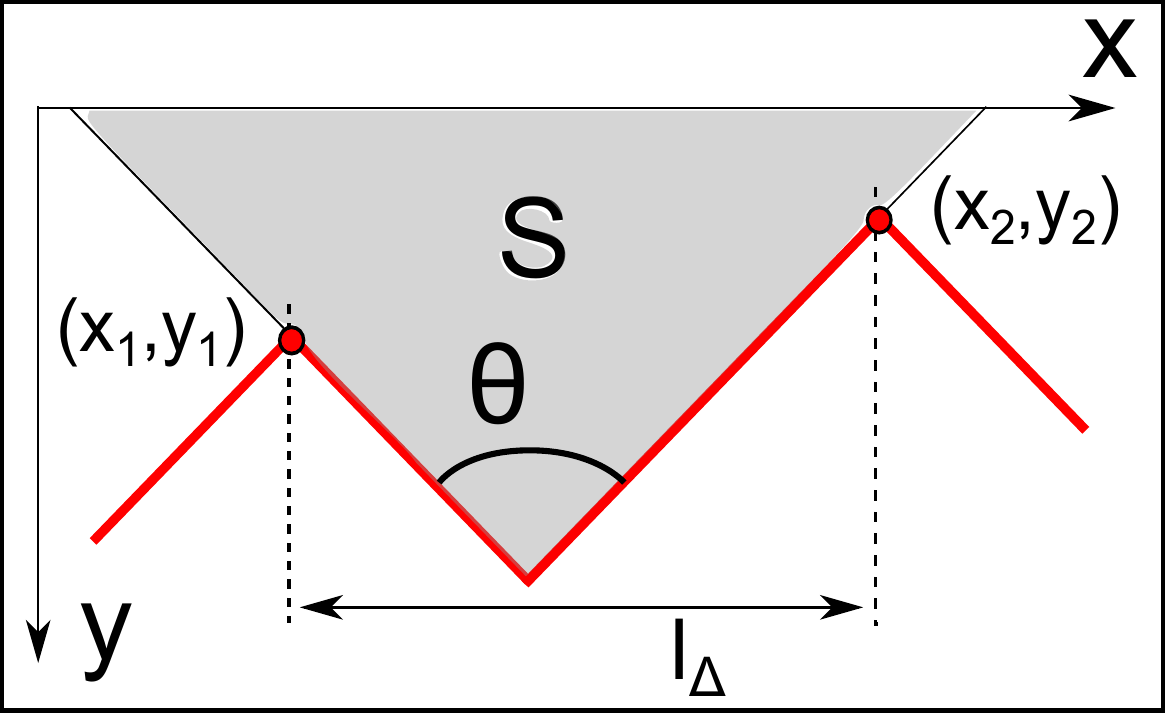}}
\centering
\end{overpic}	
\caption{Distribution of $l_{\Delta}$ for the stylized model with parameters $\lambda =0.5$ and $\theta = \pi/2$. The system size is $L=100$ and the simulation is performed over $N=3000$ samples. The dashed line corresponds to the asymptotic prediction of Eq.(\ref{interspace}). Inset: interface pinned between two adjacent stagnant patches of coordinates $\bf{x_1}$ and $\bf{x_2}$.}
\label{fig:pdfinterspace}
\end{figure}

To get a better grasp on the front roughness for $\overline{U} \lesssim U_{SD}$, we compute the distance $l_{\Delta}$ between two adjacent pinning sites. A scaling argument (that can be easily extended to various geometries) extracts the main dependence in $\lambda$ and $\theta$ of the typical distance between stagnant patches: let's assume the interface pinned at some site and consider its right part (see Inset of Fig.\ref{fig:pdfinterspace}); the probability of having another pinning is important when the area $S \sim l_{\Delta}^2/\tan(\theta/2) \sim \lambda ^{-1}$, leading to:
\begin{align}
\label{scalingarg}
l_{\Delta} \sim \sqrt{\frac{\tan{\theta/2}}{\lambda}} = \ell_{typ}
\end{align}
It turns out that it is possible to compute the whole probability distribution $\rho(l_{\Delta})$ in the \textit{sawtooth} geometry. It obeys:
\begin{align}
\rho(l_{\Delta}) &= \int_{\mathcal{D}} P(\textbf{x}_{12}) \delta (l_{\Delta} - |x_2-x_1|) \\
\nonumber
\mathcal{D} = \lbrace 0<x_i& <L , 0<y_i < \min \left(\frac{x_i}{\tan \theta /2}, \frac{L-x_i}{\tan \theta /2} \right)\rbrace
\end{align}
with $i \in \lbrace 1,2 \rbrace$. $\mathcal{D}$ simply parametrizes the area of the interface in the depinned state. $P(\textbf{x}_{12})$ is the probability that the interface is pinned in $ \textbf{x}_1=(x_1,y_1)$ and $\textbf{x}_2=(x_2,y_2)$, with no other nucleation in between:
\begin{align}
\nonumber
P(\textbf{x}_{12})&d\textbf{x}_1 d\textbf{x}_2 = \lambda^2 d \textbf{x}_1 d \textbf{x}_2 e^{- \lambda S(x_1,x_2,y_1,y_2)} \\
\nonumber
S(x_1,x_2,y_1,y_2)&= \frac{\tan(\theta/2)}{4}\left( y_1+y_2 + \frac{x_2-x_1}{\tan(\theta/2)})\right)^2 \\
&H \left(|y_2-y_1| < |x_2-x_1|/\tan (\theta /2) \right)   
\end{align}
$S$ being the triangular area depicted in the inset of Fig.\ref{fig:pdfinterspace} and $H$ a Heaviside function. Integration over the variables under the constraint that $l_{\Delta}=|x_2-x_1|$ leads, in the limit $L \to \infty$, to:
\begin{align}
\rho &(l_{\Delta}) = \frac{1}{\ell_{typ}} \hat{\rho} (r) \text{  with  } r = l_{\Delta} /\ell_{typ} \\
\nonumber
\hat{\rho}(r)=&\frac{2}{\sqrt{\pi}}\Bigg(2 \left(e^{-\frac{r^2}{4}}-e^{-r^2}\right)+  \\
  & 
   \sqrt{\pi}r \left(\text{erf}\left(\frac{r}{2}\right)-2
   \text{erf}(r)+1\right) \Bigg)
\label{interspace}
\end{align}
The maximum of $\hat{\rho}$ is of order $1$, recovering the scaling argument given in Eq.\ref{scalingarg}, and an excellent agreement with the stylized model is observed (see Fig.\ref{fig:pdfinterspace}). This distribution gives full information about the fluctuations of the static front in the porous media and allows for example to compute its lateral extension through $\overline{\Delta H} \sim l_{\Delta}/(2\tan(\theta/2))$. Finer details about the statistical properties of the interface can be useful, for example to study fluctuations of the critical currents of strongly pinned vortex in superconductors \citep{koshelev_theory_2011}.

In this Letter, we presented a general model of pinning for interfaces in random media, when the pinning regime is strong and only relies on a finite number of sites. This in particular makes an approach through Poisson processes possible, allowing at the same time efficient numerical simulations and analytical results on the statistical properties of the interface. The essential experimental picture \citep{atis_autocatalytic_2013} is reproduced and we identified a clear criterium that allows to discriminate between the possible scenarios shown in Fig.\ref{numericsphase}. Supported by excellent agreement with {\it ab initio} simulations used to model the experiments \citep{saha_phase_2013}, this validates the hypothesis that the depinning transition is controlled by a limited number of events, randomly spread over the medium. 

The above model assumes the interfaces in its final state. However, strong pinning phenomenons often exhibit avalanches during transient phases, where some stagnant patches temporarily pin the interface before getting suddenly depleted. The temporal critical properties of those systems are not well understood. As a perspective, it would hence be interesting to extend the present work to transient states by introducing random life time of the nucleation events.
We are grateful to acknowledge Severine Atis, Pierre Le Doussal and Dominique Salin for useful discussions.

\appendix
\section{Generation of random fields}\label{random}

In this paper we solve Eq.(2) of the main text for a random permeability field $K(\vec r)$ correlated on a distance $\ell _d$. We denote the cumulative distribution by cdf. Two distributions were employed: 
\begin{itemize}
\item
The log-normal, of cdf $\Pi (K) = \frac{1}{2} \left[ 1 + {\rm erf}(\frac{ \log K-\bar K}{\sqrt{2}\sigma }) \right]$ where $\bar K$ is the log-scale and $\sigma$ the shape.
\item 
The Weibull, of cdf $\Pi (K) = 1- e^{- (K/\lambda)^{\delta}}$ where $\lambda$ is the scale and $\delta$ controls the decay of $\Pi(K)$ close to zero : $\Pi(K) \sim (\frac{K}{\lambda})^{\delta} $
\end{itemize}
This random field is generated as follows: 

\textit{Step $1$}. We first generate a Gaussian random field $Z(\vec r)$ correlated on a scale $\ell_d$ using the Fast Fourier Transform method (FFT): 
\begin{align}
Z(\vec r)= \sum_{\vec k \neq 0} Z_{\vec k} e^{i \vec k \cdot \vec r} \\
\label{correl}
\overline{Z_{\vec k} Z_{{\vec k}'}} = \delta _{\vec k,- \vec k'} e^{- k^2/k_d^2} 
\end{align}
where $\vec k=\frac{2 \pi}{L} \vec n$, $\vec n \in \left\{-L/2+1,\ldots,L/2-1,L/2  \right\}^2$, $k_d= \pi / \ell _d$ is the cut-off for the correlation in the momentum space and $Z_{\vec k}$ a vector of uncorrelated normal numbers. Therefore, the mean $\overline{Z}(\vec r) =0$ and the variance $\sigma ^2 _Z = \sum _{\vec k} e^{- k^2 /k_d ^2} $. 

\textit{Step $2$}.
We transform the Gaussian statistics into the families of distributions detailed above by the Inverse transform sampling method. We denote $F_G(x)$ the cdf of the normal distribution of mean $\mu$ and variance $\sigma^2$:
\begin{equation}
F_G (x)= \frac{1}{2} \left[1 +  {\rm erf} \left( \frac{x-\mu}{\sqrt{2} \sigma } \right) \right]
\end{equation}
Each component of the field $F_G(Z(\vec r))$ is then uniformly distribution over $[0,1]$. It is possible to revert to a general random field $X$ of cdf $F_X$ by applying $F^{-1}_X$. For the families considered in the main paper, this leads to:
\begin{itemize}
 \item $K(\vec r) = \exp(Z(\vec r))$  for the log-normal distribution.
 \item $K(\vec r) =   -\lambda \left( \log(1/2(1 - {\rm erf}(\frac{Z(\vec r) - \mu}{\sqrt{2} \sigma } ) ) \right)^{1/\delta}  $, for the Weibull distribution.
\end{itemize}
Note that the correlations remain short range over a length $\ell _d$, although they are not of the form given in Eq.(\ref{correl}).

\bibliography{bibporous} 

\begin{thebibliography}{10}

\bibitem{adler_chemotaxis_1966}
Julius Adler.
\newblock Chemotaxis in bacteria.
\newblock {\em Science}, 153(3737):708--716, December 1966.
\newblock {PMID:} 4957395.

\bibitem{boyd2003physics}
T.J.M. Boyd and J.J. Sanderson.
\newblock {\em The Physics of Plasmas}.
\newblock Cambridge University Press, Cambridge, 2003.

\bibitem{jarosinski_combustion_2009}
Jozef Jarosinski and Bernard Veyssiere.
\newblock {\em Combustion Phenomena: Selected Mechanisms of Flame Formation,
  Propagation and Extinction}.
\newblock {CRC} Press, February 2009.

\bibitem{fort_progress_2008}
Joaquim Fort and Toni Pujol.
\newblock Progress in front propagation research.
\newblock {\em Rep. Prog. Phys.}, 71(8):086001, August 2008.

\bibitem{fisher_wave_1937}
R.~A. Fisher.
\newblock The wave of advance of advantageous genes.
\newblock {\em Annals of Eugenics}, 7(4):355–369, 1937.

\bibitem{edwards_poiseuille_2002}
Boyd~F. Edwards.
\newblock Poiseuille advection of chemical reaction fronts.
\newblock {\em Phys. Rev. Lett.}, 89(10):104501, August 2002.

\bibitem{schwartz_chemical_2008}
M.~E. Schwartz and T.~H. Solomon.
\newblock Chemical reaction fronts in ordered and disordered cellular flows
  with opposing winds.
\newblock {\em Phys. Rev. Lett.}, 100(2):028302, January 2008.

\bibitem{xin_front_2000}
Jack Xin.
\newblock Front propagation in heterogeneous media.
\newblock {\em {SIAM} Rev.}, 42(2):161–230, June 2000.

\bibitem{koptyug_advection_2008}
Igor~V. Koptyug, Vladimir~V. Zhivonitko, and Renad~Z. Sagdeev.
\newblock Advection of chemical reaction fronts in a porous medium.
\newblock {\em The Journal of Physical Chemistry B}, 112(4):1170--1176, January
  2008.

\bibitem{korzhavin_dynamics_1997}
V.~A.~Bunev A.~A.~Korzhavin.
\newblock Dynamics of gaseous combustion in closed systems with an inert porous
  medium.
\newblock {\em Combustion and Flame}, (4):507--520, 1997.

\bibitem{kuo_theory_1973}
{K.K.} Kuo, R.~Vichnevetsky, and M.~Summerfield.
\newblock Theory of flame front propagation in porous propellant charges under
  confinement.
\newblock {\em {AIAA} Journal}, 11(4):444--451, April 1973.

\bibitem{atis_self-sustained_2012}
Severine Atis, Sandeep Saha, Harold Auradou, Dominique Salin, and Laurent
  Talon.
\newblock Self-sustained reaction fronts in porous media.
\newblock {\em {arXiv:1210.3518}}, October 2012.
\newblock {PhysRevLett.110.148301} (2013).

\bibitem{atis_autocatalytic_2013}
Severine Atis, Sandeep Saha, Harold Auradou, Dominique Salin, and Laurent
  Talon.
\newblock Autocatalytic reaction fronts inside a porous medium of glass
  spheres.
\newblock {\em Phys. Rev. Lett.}, 110(14):148301, April 2013.

\bibitem{yang_nanostructured_1997}
Peidong Yang and Charles~M. Lieber.
\newblock Nanostructured high-temperature superconductors: Creation of
  strong-pinning columnar defects in nanorod/superconductor composites.
\newblock {\em Journal of Materials Research}, 12(11):2981--2996, 1997.

\bibitem{maiorov_synergetic_2009}
B.~Maiorov, S.~A. Baily, H.~Zhou, O.~Ugurlu, J.~A. Kennison, P.~C. Dowden,
  T.~G. Holesinger, S.~R. Foltyn, and L.~Civale.
\newblock Synergetic combination of different types of defect to optimize
  pinning landscape using {BaZrO3-doped} {YBa2Cu3O7}.
\newblock {\em Nat Mater}, 8(5):398--404, May 2009.

\bibitem{koshelev_theory_2011}
A.~E. Koshelev and A.~B. Kolton.
\newblock Theory and simulations on strong pinning of vortex lines by
  nanoparticles.
\newblock {\em Phys. Rev. B}, 84(10):104528, September 2011.

\bibitem{dalmas_crack_2009}
D.~Dalmas, E.~Barthel, and D.~Vandembroucq.
\newblock Crack front pinning by design in planar heterogeneous interfaces.
\newblock {\em Journal of the Mechanics and Physics of Solids}, 57(3):446--457,
  March 2009.

\bibitem{jarrige10a}
N.~Jarrige, I.~Bou~Malham, J.~Martin, N.~Rakotomalala, D.~Salin, and L.~Talon.
\newblock Numerical simulations of a buoyant autocatalytic reaction front in
  tilted hele-shaw cells.
\newblock {\em Phys. Rev. E}, 81:066311, 2010.

\bibitem{matheron}
G.~Matheron.
\newblock {\em {\'E}l{\'e}ments pour une th{\'e}orie des milieux poreux}.
\newblock Masson et Cie, Paris, 1967.

\bibitem{talon03}
L.~Talon, J.~Martin, N.~Rakotomalala, D.~Salin, and Y.C. Yortsos.
\newblock {L}attice {B}{G}{K} simulations of macrodispersion in heterogeneous
  porous media.
\newblock {\em Water Resour. Res.}, 39:1135--1142, 2003.

\bibitem{ginzburg08c}
I.~Ginzburg.
\newblock Consistent lattice boltzmann schemes for the brinkman model of porous
  flow and infinite chapman-enskog expansion.
\newblock {\em Physical Review E (Statistical, Nonlinear, and Soft Matter
  Physics)}, 77:066704, 2008.

\bibitem{lecontemartinsalin}
M.~Leconte, J.~Martin, N.~Rakotomalala, and D.~Salin.
\newblock Pattern of reaction diffusion fronts in laminar flows.
\newblock {\em Phys. Rev. Lett.}, 90:128302, 2003.

\bibitem{williamsconb}
F.~Williams.
\newblock {\em Combustion theory, 2nd Ed.}
\newblock Benjamin/Cummings, New York, 1985.

\bibitem{saha_phase_2013}
Sandeep Saha, Severine Atis, Dominique Salin, and Laurent Talon.
\newblock Phase diagram of sustained wave fronts opposing the flow in
  disordered porous media.
\newblock {\em {EPL}}, 101(3):38003, February 2013.

\end{thebibliography}

\end{document}